\documentclass[11pt,a4paper]{article}
\pdfoutput=1
\usepackage{jheppub}
\usepackage{tikz}
\usetikzlibrary{intersections, calc, arrows.meta, bending}
\usepackage{subcaption}
\DeclareMathOperator{\Tr}{Tr}
\DeclareMathOperator{\tr}{tr}

\newcommand{\ri}{\mathrm{i}}
\renewcommand{\th}{\theta}

\newcommand{\cob}{\delta}

\newcommand{\vth}{\vartheta}

\newcommand{\hf}{\frac{1}{2}}

\newcommand{\del}{\partial}

\newcommand{\lap}{\Delta}
\newcommand{\bra}{\langle}
\newcommand{\ket}{\rangle}
\newcommand{\la}{\lambda}

\newcommand{\bt}{\beta}

\newcommand{\al}{\alpha}
\newcommand{\om}{\omega}

\newcommand{\rt}[1]{\sqrt{#1}}
\newcommand{\cO}{\mathcal{O}}
\newcommand{\cZ}{\mathcal{Z}}

\newcommand{\cN}{\mathcal{N}}

\newcommand{\tq}{\tilde{q}}

\makeatletter
\gdef\@fpheader{}
\makeatother
\begin{document}
\title{Solvable limit of ETH matrix model for double-scaled SYK}

\author[a]{Kazumi Okuyama}
\author[b]{and Takao Suyama}

\affiliation[a]{Department of Physics, 
Shinshu University, 3-1-1 Asahi, Matsumoto 390-8621, Japan}
\affiliation[b]{High Energy Accelerator Research Organization (KEK),
Oho 1-1, Tsukuba, Ibaraki 305-0801, Japan}

\emailAdd{kazumi@azusa.shinshu-u.ac.jp,tsuyama@post.kek.jp}

\abstract{
We study the two-matrix model for double-scaled SYK model, called ETH matrix model
introduced by Jafferis et al [arXiv:2209.02131].
If we set the parameters $q_A,q_B$ of this model to zero,
the potential of this two-matrix model is given by the Gaussian
terms and the $q$-commutator squared interaction.
We find that this model is solvable in the large $N$ limit
and we explicitly construct the planar one- and two-point function
of resolvents
in terms of elliptic functions. 
}

\maketitle

\section{Introduction}
Sachdev-Ye-Kitaev (SYK) model is a very useful toy model for the
study of quantum gravity  
\cite{Sachdev1993,Kitaev1,Kitaev2,Polchinski:2016xgd,Maldacena:2016hyu}.
At low energy, SYK model is described by the Schwarzian mode
which is holographically dual to the Jackiw-Teitelboim (JT) gravity
\cite{Jackiw:1984je,Teitelboim:1983ux}.
In the seminal paper \cite{Saad:2019lba}, 
it was found that JT gravity is equivalent to a random matrix model
in the double scaling limit.

One can go beyond the low energy Schwarzian
approximation by taking a certain double scaling limit
of the SYK model \cite{Cotler:2016fpe,Berkooz:2018jqr}, 
which we will call double-scaled SYK (DSSYK) in this paper.
As shown in \cite{Berkooz:2018jqr}, DSSYK is exactly solvable in the
large $N$ limit 
using the technique of the chord diagram and the transfer matrix.
One can also compute the correlators of matter operator $\cO$
in DSSYK using the chord diagram, assuming that
the coefficient $K$ in $\cO$ and the coefficient $J$
in the Hamiltonian $H$ are independent Gaussian-random variables 
\cite{Berkooz:2018jqr}.
Thus DSSYK is capable of describing
the holographic dual of (a certain $q$-deformation of) JT gravity coupled to
a propagating matter field.

As shown in \cite{Jafferis:2022wez},
the correlators of matter operator $\cO$ in DSSYK 
is written as a two-matrix model with a single trace potential
\begin{equation}
\begin{aligned}
 \cZ=\int dA dB e^{-N\Tr V(A,B)},
\end{aligned}
\label{eq:cZAB} 
\end{equation}
which is called ETH matrix model in \cite{Jafferis:2022wez}.\footnote{ETH stands for
the eigenstate thermalization hypothesis \cite{deutsch1991quantum,srednicki1994chaos}.} Here two matrices $A$
and $B$ correspond to $H$ and $\cO$, respectively.
One can compute the potential $V(A,B)$ as a series
expansion in the parameters $q_A,q_B$ of DSSYK
by matching the correlators of matter operator $\cO$ at the disk level
(see \eqref{eq:scaling} and \eqref{eq:qB-def} for the definition of 
the parameters $q_A,q_B$).

If we ignore the effect of matter matrix $B$,
the potential for the matrix $A$ can be computed in a closed form
\cite{Jafferis:2022wez}.
This one-matrix model for DSSYK reduces to the matrix model of 
JT gravity \cite{Saad:2019lba} by zooming in on 
the edge of the eigenvalue distribution and 
taking the double scaling limit.
Interestingly, the one-matrix model for DSSYK
makes sense  in the ordinary
't Hooft expansion without taking the double scaling limit.\footnote{
This is similar in spirit to the open/closed duality of Gaussian matrix model
advocated by Gopakumar and collaborators \cite{Gopakumar:2011ev,Gopakumar:2012ny,Gopakumar:2022djw}.
}
As shown in \cite{Okuyama:2023kdo}, the correlators of one-matrix model for DSSYK
in this 't Hooft expansion can be decomposed into the 
trumpet and the volume of moduli space, 
in a completely parallel way as the JT gravity matrix model. 
One important difference from JT gravity is that the bulk geodesic length 
becomes discrete in DSSYK \cite{Jafferis:2022wez,Lin:2022rbf,Okuyama:2023kdo,Okuyama:2023byh}.

We are interested in the effect of matter field in the bulk
gravity theory.
In particular, we would like to understand the loop
correction to the wormhole amplitude coming from the matter loop
running around the neck of the wormhole.
In JT gravity, this loop correction is UV divergent
due to the contribution from long, thin wormhole 
\cite{Saad:2019lba,Penington:2019kki}.
We expect that DSSYK gives a UV completion of 
JT gravity coupled to a matter field
and the wormhole amplitude computed from the two-matrix model
\eqref{eq:cZAB} is free of divergence.
However, the potential $V(A,B)$ for DSSYK is very complicated and
it seems hopeless to solve
the two-matrix model \eqref{eq:cZAB} even in the planar limit.

It turns out that when $q_A=q_B=0$, the 
two-matrix model \eqref{eq:cZAB} for DSSYK drastically simplifies and
becomes solvable in the large $N$ limit.
In this paper, we compute the one- and two-point functions of
resolvent $\Tr\frac{1}{\la-A}$ in this solvable limit of two-matrix model.
In this case, the potential $V(A,B)$ is given by the Gaussian
terms together with the $q$-commutator squared interaction (see \eqref{eq:VAB}).
It turns out that the matrix $B$ can be integrated out in this case,
and the partition function \eqref{eq:cZAB} is written as the eigenvalue integral
of the matrix $A$.
We find that the saddle-point equation for
this eigenvalue integral is very similar to
the one appeared in the Dijkgraaf-Vafa matrix model 
of $\bt$-deformed $\cN=4$ super Yang-Mills 
\cite{Dijkgraaf:2002dh,Dorey:2002tj,Dorey:2002pq}
and the six vertex model on a random lattice \cite{Kostov:1999qx}.
As a consequence, the planar resolvent
of our two-matrix model
can be written in terms of an elliptic function,
in a similar manner as \cite{Dijkgraaf:2002dh,Dorey:2002tj,Dorey:2002pq,Kostov:1999qx}.
We also find that the two-point function
of resolvents is given by the Bergman kernel on a torus.
In this computation, it is useful to introduce the 't Hooft parameter $S$
as in \eqref{eq:VAB-S}.
The original matrix model potential \eqref{eq:VAB}
corresponds to the $S=1$ case.
We find that the planar solution of resolvent
behaves differently for $S=1$ and $S<1$.

This paper is organized as follows. In section \ref{sec:review},
we review the ETH matrix model of DSSYK defined in \cite{Jafferis:2022wez}
and rewrite it as the eigenvalue integral 
when $q_A=q_B=0$.
In section \ref{sec:one}, we write down the large $N$
saddle-point equation for the resolvent
and find a solution which respects the $\mathbb{Z}_2$ symmetry 
$A\to-A$ of the model.
In section \ref{sec:two}, we compute the two-point function of resolvents
in the large $N$ limit.
We find that the two-point function is given by the Bergman kernel on a torus.
Finally, we conclude in section \ref{sec:conclusion}
with some discussion on future problems.
In appendix \ref{app:moment} we compute the even part of the moments
of two-point function for $S<1$.

\section{Review of ETH matrix model}\label{sec:review}
In this section, we review the ETH matrix model of DSSYK introduced 
in \cite{Jafferis:2022wez}. 

\subsection{Review of DSSYK}
SYK model is defined by the Hamiltonian for 
$M$ Majorana fermions $\psi_i~(i=1,\cdots,M)$
obeying $\{\psi_i,\psi_j\}=2\cob_{i,j}$
with all-to-all $p$-body interaction
\begin{equation}
\begin{aligned}
 H=\ri^{p/2}\sum_{1\leq i_1<\cdots<i_p\leq M}
J_{i_1\cdots i_p}\psi_{i_1}\cdots\psi_{i_p},
\end{aligned} 
\end{equation}
where $J_{i_1\cdots i_p}$ is a random coupling drawn from the Gaussian distribution
with the mean and the variance given by
\begin{equation}
\begin{aligned}
 \bra J_{i_1\cdots i_p}\ket_J=0,\quad \bra J_{i_1\cdots i_p}^2\ket_J=\binom{M}{p}^{-1}.
\end{aligned} 
\end{equation}
DSSYK is defined by the scaling limit
\begin{equation}
\begin{aligned}
 M,p\to\infty\quad\text{with}\quad q_A=e^{-\frac{2p^2}{M}}:\text{fixed}.
\end{aligned} 
\label{eq:scaling}
\end{equation}
As shown in \cite{Berkooz:2018jqr}, the ensemble average of the moment $\tr H^k$ 
reduces to a counting problem
of the intersection number of chord diagrams
\begin{equation}
\begin{aligned}
 \bra \tr H^k\ket_J=\sum_{\text{chord diagrams}}q_A^{\#(\text{intersections})}.
\end{aligned} 
\end{equation}
Here, $\tr$ in $\tr H^k$ refers to the trace over the Fock space of Majorana fermions.
Using the technique of the transfer matrix,
the disk amplitude $\bra \tr e^{-\bt H}\ket_J$ of DSSYK 
is explicitly evaluated as \cite{Berkooz:2018jqr}
\begin{equation}
\begin{aligned}
 \bra \tr e^{-\bt H}\ket_J=\int_0^\pi\frac{d\th}{2\pi}\mu(\th)
e^{-\bt E(\th)},
\end{aligned} 
\label{eq:disk-DSSYK}
\end{equation}
where $\mu(\th)$ and $E(\th)$ are given by
\begin{equation}
\begin{aligned}
\mu(\th)=(q_A,e^{\pm2\ri\th};q_A)_\infty,\qquad
 E(\th)=\frac{2\cos\th}{\rt{1-q_A}}.
\end{aligned} 
\label{eq:E-th}
\end{equation}
The $q$-Pochhammer symbol is defined by
\begin{equation}
\begin{aligned}
 (a;q)_\infty=\prod_{k=0}^\infty (1-aq^k),
\end{aligned} 
\end{equation}
and the measure factor in \eqref{eq:E-th}
is a shorthand of
\begin{equation}
\begin{aligned}
 (q_A,e^{\pm2\ri\th};q_A)_\infty&=(q_A;q_A)_\infty (e^{2\ri\th};q_A)_\infty
(e^{-2\ri\th};q_A)_\infty.
\end{aligned} 
\end{equation}

One can also introduce
a matter field in the bulk which is dual to an
operator in DSSYK. One simple example 
is the length $s$ strings of Majorana fermions
\begin{equation}
\begin{aligned}
 \cO=\ri^{s/2}\sum_{1\leq i_1<\cdots< i_s\leq M}K_{i_1\cdots i_s}\psi_{i_1}\cdots\psi_{i_s}
\end{aligned} 
\label{eq:matter-M}
\end{equation}
with Gaussian random coefficients $K_{i_1\cdots i_s}$
which is drawn independently from the random coupling
$J_{i_1\cdots i_p}$ in the SYK Hamiltonian.
The effect of this operator can be made finite by taking the limit
$M,s,p\to\infty$ with the following combinations held fixed:
\begin{equation}
\begin{aligned}
 \tq=e^{-\frac{2sp}{M}},\quad q_B=e^{-\frac{2s^2}{M}}.
\end{aligned} 
\label{eq:qB-def}
\end{equation}
In this limit, the random average of the correlator
$\tr(e^{-\bt_2 H}\cO e^{-\bt_1 H}\cO)$ becomes
\begin{equation}
\begin{aligned}
\bra\tr(e^{-\bt_2 H}\cO e^{-\bt_1 H}\cO)\ket_{J,K}=\int\prod_{i=1,2}\frac{d\th_i}{2\pi}\mu(\th_i)e^{-\bt_i E(\th_i)}\frac{1}{F(\th_1,\th_2)}
\end{aligned} 
\label{eq:2pt}
\end{equation}
with
\begin{equation}
\begin{aligned}
F(\th_1,\th_2)= 
\frac{(\tq e^{\ri(\pm\th_1\pm\th_2)};q_A)_\infty}{(\tq^2;q_A)_\infty}.
\end{aligned} 
\label{eq:F12}
\end{equation}

\subsection{ETH matrix model of DSSYK}
As discussed in \cite{Jafferis:2022wez},
one can construct a two-matrix model 
for $N\times N$ matrices $A$ and $B$ which reproduces
the disk amplitude \eqref{eq:disk-DSSYK} and the matter two-point function
\eqref{eq:2pt} in the large $N$ limit.
The two matrices $A$ and $B$ correspond to $H$ and $\cO$ in DSSYK, respectively
\begin{equation}
\begin{aligned}
 A\leftrightarrow H,\quad B\leftrightarrow \cO,
\end{aligned} 
\end{equation}
and the size $N$ of the matrices $A,B$ is given by the dimension of the Hilbert space
of $M$ Majorana fermions
\begin{equation}
\begin{aligned}
 N=2^{M/2}.
\end{aligned} 
\end{equation}

As demonstrated in \cite{Jafferis:2022wez},
one can construct the potential $V(A,B)$ of two-matrix model
\eqref{eq:cZAB} 
order by order in the small $(q_A,q_B)$ expansion.
At the leading order in the small $q_B$ expansion, the potential is written as
\begin{equation}
\begin{aligned}
 \Tr V(A,B)=\sum_{a=1}^N\Bigl[V_0(\la_a)+V_{\text{ct}}(\la_a)\Bigr]+\hf \sum_{a,b=1}^N B_{ab}B_{ba}F(\la_a,\la_b)
\end{aligned} 
\label{eq:Vsum}
\end{equation}
where we diagonalized the matrix $A$ and denoted the eigenvalue of $A$
as $\{\la_a\}_{a=1,\cdots,N}$.
In the last term of \eqref{eq:Vsum}, $F(\la_a,\la_b)$ is given by
$F(\th_a,\th_b)$ in \eqref{eq:F12} with the identification
$\la_a=E(\th_a)$.
The last term of \eqref{eq:Vsum} is constructed
in such a way that the propagator of $B$ reproduces
the matter two-point function \eqref{eq:2pt} of DSSYK.
The first term $V_0(\la)$ in \eqref{eq:Vsum} is determined by requiring
that the eigenvalue density $\rho_0(\la)$ for the
matrix $A$ agrees with $\mu(\th)$ in \eqref{eq:E-th} if we ignore $B$.
The second term $V_{\text{ct}}(\la)$ in \eqref{eq:Vsum}
is the counter term introduced so that
the one-loop correction coming from integrating out $B$
is canceled and the eigenvalue density $\rho_0(\la)$
is not modified at the leading order in the large $N$ limit
\begin{equation}
\begin{aligned}
 V_{\text{ct}}(\la)=-\int d\la'\rho_0(\la')\log F(\la',\la).
\end{aligned}
\label{eq:Vct} 
\end{equation}
The explicit form of $V_0(\la)$ and $V_{\text{ct}}(\la)$
is given by \cite{Jafferis:2022wez}
\begin{equation}
\begin{aligned}
 V_0(\la)&=\sum_{n=1}^\infty (-1)^{n-1}
\frac{q_A^{\hf n^2}}{n}(q_A^{\hf n}+q_A^{-\hf n})
T_{2n}\left(\frac{\rt{1-q_A}}{2}\la\right),\\
V_{\text{ct}}(\la)&=\sum_{n=1}^\infty (-1)^{n}
\frac{\tq^{2n}}{1-q_A^{2n}}\frac{q_A^{\hf n^2}}{n}(q_A^{\hf n}+q_A^{-\hf n})
T_{2n}\left(\frac{\rt{1-q_A}}{2}\la\right),
\end{aligned} 
\label{eq:V0-Vct}
\end{equation}
where $T_n(\cos\th)=\cos(n\th)$ denotes the Chebyshev polynomial of the first kind.

For a general value of $\tq,q_A$ and $q_B$,
it seems hopeless to solve the two-matrix model \eqref{eq:cZAB}
even in the planar limit.
However, it turns out that the two-matrix model
becomes solvable in the limit\footnote{Physically, this limit corresponds to keeping
only the intersections between $H$-chord and $\cO$-chord and discarding
the chord diagrams with $H$-$H$ and $\cO$-$\cO$ intersections.}
\begin{equation}
\begin{aligned}
 q_A,q_B\to0,\quad \tq=\text{finite}.
\end{aligned} 
\label{eq:qlimit}
\end{equation}
In this limit \eqref{eq:qlimit}, the matrix model potential becomes
(see eq.(8.52) in \cite{Jafferis:2022wez})
\begin{equation}
\begin{aligned}
 \Tr V(A,B)&=\Tr\left[\hf(1-\tq^2)(A^2+B^2)+\frac{\tq^2}{1-\tq^2}A^2B^2
-\frac{\tq(1+\tq^2)}{2(1-\tq^2)}ABAB\right]\\
&=
\hf\Tr\left[(1-\tq^2)(A^2+B^2)
-\frac{\tq^2}{1-\tq^2}
\Bigl([A,B]_{\tq}\Bigr)^2\right],
\end{aligned} 
\label{eq:VAB}
\end{equation}
where we introduced the $q$-deformed commutator $[A,B]_{\tq}$ as
\begin{equation}
\begin{aligned}
 [A,B]_{\tq}=\tq^{\hf }AB-\tq^{-\hf}BA.
\end{aligned} 
\end{equation}
This potential \eqref{eq:VAB} is obtained from the $q_A\to0$ limit of 
\eqref{eq:Vsum} since we have already set $q_B=0$ in \eqref{eq:Vsum}.
Note that our two-matrix model with the potential
\eqref{eq:VAB} is reminiscent of the models appeared in
the context of 
the Dijkgraaf-Vafa model for the $\cN=4$ super Yang-Mills with 
$\bt$-deformation \cite{Dijkgraaf:2002dh,Dorey:2002tj,Dorey:2002pq,Rossi:2009mn,Mansson:2003dm}
and the six vertex model on a random lattice  
\cite{Kostov:1999qx}. \footnote{See also \cite{Hoppe:1999xg,Zakany:2018dio,Eynard:1995zv,hoppe1989quantum} for matrix models related to our model.}

We are interested in the disk and the cylinder amplitude of our model
\eqref{eq:VAB}
in the large $N$ limit
\begin{equation}
\begin{aligned}
 \Bigl\bra \Tr e^{-\bt A}\Bigr\ket&=N\int d\la\rho(\la)e^{-\bt\la},\\
\Bigl\bra \Tr e^{-\bt_1 A}\Tr e^{-\bt_2 A}\Bigr\ket_{\text{conn}}&=\int d\la_1d\la_2\rho(\la_1,\la_2)e^{-\bt_1\la_1-\bt_2\la_2},
\end{aligned} 
\end{equation}
which are obtained once we know the one- and the two-point
function of the resolvent
\begin{equation}
\begin{aligned}
 R(\la)=\Tr\frac{1}{\la-A}.
\end{aligned} 
\label{eq:resolvent}
\end{equation}
In order to study this two-matrix model, 
it is convenient to introduce the 't Hooft parameter $S$
\footnote{
We follow the convention of Dijkgraaf-Vafa model \cite{Dijkgraaf:2002dh}
to denote the 't Hooft parameter as $S$.}
in the potential \eqref{eq:VAB}
\begin{equation}
\begin{aligned}
 \Tr V(A,B)=\hf\Tr\left[(1-\tq^2)\Bigl(\frac{A^2}{S}+B^2\Bigr)
-\frac{\tq^2}{1-\tq^2}
\Bigl([A,B]_{\tq}\Bigr)^2\right].
\end{aligned} 
\label{eq:VAB-S}
\end{equation}
As shown in \cite{Jafferis:2022wez},
this modification of the potential naturally arises when we turn on $q_A\ne0$
\begin{equation}
\begin{aligned}
 S=1+\cO(q_A).
\end{aligned} 
\end{equation}
In the rest of this paper, we will study the large $N$ limit of
the two-matrix model with the potential in \eqref{eq:VAB-S}.
As we will see below, $S=1$ is a somewhat singular limit and setting $S\ne1$
gives a natural regularization
of the model.

\section{One-point function of resolvent}\label{sec:one}
In this section, we study the one-point function of resolvent
\eqref{eq:resolvent} in the large $N$ planar limit.

\subsection{Saddle-point equation}
One can easily see that $B$ 
can be integrated out in our two-matrix model since $B$ appears quadratically in the potential
$V(A,B)$ in \eqref{eq:VAB-S}.
After integrating out $B$, the 
partition function $\cZ$
of two-matrix model 
\eqref{eq:cZAB} is written as the integral over 
the eigenvalues $\{\la_a\}_{a=1,\cdots, N}$ of $A$
\begin{equation}
\begin{aligned}
 \cZ&=\int\prod_{a=1}^N d\la_a e^{-\frac{N(1-\tq^2)\la_a^2}{2S}}
\frac{\prod_{a<b}(\la_a-\la_b)^2}{\prod_{a,b}\rt{(1-\tq^2)^2+\tq^2
(\tq^\hf\la_a-\tq^{-\hf}\la_b)
(\tq^{-\hf}\la_a-\tq^{\hf}\la_b)}}\\
&=\int\prod_{a=1}^N \frac{d\la_a e^{-\frac{N(1-\tq^2)\la_a^2}{2S}}}
{\rt{(1+\tq)^2-\tq\la_a^2}}
\prod_{a<b}\frac{(\la_a-\la_b)^2}{(1-\tq^2)^2+\tq^2
(\la_a-\tq^{-1}\la_b)
(\la_a-\tq\la_b)}.
\end{aligned} 
\label{eq:cZ-la}
\end{equation}
Here we ignored the overall normalization constant.

We would like to analyze the saddle-point equation
for the integral \eqref{eq:cZ-la} in the large $N$ limit.
To this end, it is convenient to
make a change of variables
\begin{equation}
\begin{aligned}
 \la_a=2\cosh z_a.
\end{aligned}
\label{eq:la-a} 
\end{equation} 
Then $\cZ$ in \eqref{eq:cZ-la} is written as
\begin{equation}
\begin{aligned}
 \cZ=\int \prod_{a=1}^N\frac{dz_a \sinh z_a
e^{-\frac{2 N}{S}(1-e^{-2\lap})\cosh^2 z_a}}{\rt{\cosh^2\lap-
\cosh^2z_a}}
\prod_{a<b}\frac{(\cosh z_a-\cosh z_b)^2}{\bigl(\cosh(z_a+\lap)-\cosh z_b\bigr)
\bigl(\cosh(z_a-\lap)-\cosh z_b\bigr)},
\end{aligned} 
\label{eq:cZ-z}
\end{equation}
where $\lap$ is related to $\tq$ by
\begin{equation}
\begin{aligned}
 \tq=e^{-\lap}.
\end{aligned} 
\end{equation}
The large $N$ saddle-point equation 
for the integral \eqref{eq:cZ-z} is given by
\begin{equation}
\begin{aligned}
 &4(1-e^{-2\lap})\cosh z_a\sinh z_a\\
=&\frac{S}{N}
\sum_{b\ne a}\left[\frac{2\sinh z_a}{\cosh z_a-\cosh z_b}-
\frac{\sinh(z_a+\lap)}{\cosh(z_a+\lap)-\cosh z_b}-
\frac{\sinh(z_a-\lap)}{\cosh(z_a-\lap)-\cosh z_b}
\right].
\end{aligned} 
\label{eq:saddle}
\end{equation}
Note that the factor $\frac{\sinh z_a}{\rt{\cosh^2\lap-\cosh^2z_a}}$
in the integration measure of \eqref{eq:cZ-z} 
can be ignored in the saddle-point equation
\eqref{eq:saddle} since its effect is sub-leading in the large $N$
limit.

Introducing the one-point function
$\om(z)$ by
\begin{equation}
\begin{aligned}
 \om(z)
=\frac{1}{N}\sum_{b=1}^N\frac{\sinh z}{\cosh z-\cosh z_b},
\end{aligned} 
\label{eq:om-z}
\end{equation}
the saddle-point equation \eqref{eq:saddle} is written as
\begin{equation}
\begin{aligned}
 2(1-e^{-2\lap})\sinh(2z)=S\bigl[2\om(z)-\om(z+\lap)-\om(z-\lap)\bigr].
\end{aligned} 
\label{eq:eom}
\end{equation}
If we further define $G(z)$ by
\begin{equation}
\begin{aligned}
 G(z)=2e^{-\lap}\cosh(2z)+S\bigl[\om(z+\lap/2)-\om(z-\lap/2)\bigr],
\end{aligned} 
\label{eq:def-Gz}
\end{equation} 
the saddle-point equation \eqref{eq:eom} 
implies that $G(z)$ satisfies
\begin{equation}
\begin{aligned}
 G(z+\lap/2)=G(z-\lap/2).
\end{aligned} 
\label{eq:G-eqz}
\end{equation}
This is similar to the relation appeared in \cite{Dorey:2002pq,Kostov:1999qx} and 
hence 
$G(z)$ is solved by an elliptic function, as we will see below.

\subsection{Symmetry of $\om(X)$}
We would like to find 
a solution $G(z)$ to the equation \eqref{eq:G-eqz}
which respects the symmetry of our two-matrix model \eqref{eq:VAB-S}.
For this purpose, it is convenient to introduce the 
variable $X$ as
\begin{equation}
\begin{aligned}
 X=e^z
\end{aligned} 
\label{eq:X-def}
\end{equation}
and denote $\om(z)$ and $G(z)$ as $\om(X)$ and $G(X)$, respectively.
From \eqref{eq:la-a}, the eigenvalue $\la$ and $X$ are related by
\begin{equation}
\begin{aligned}
 \la=X+X^{-1}.
\end{aligned} 
\label{eq:la-X}
\end{equation}
This is known as the Joukowsky map.  
Then $G(X)$ in \eqref{eq:def-Gz} becomes
\begin{equation}
\begin{aligned}
 G(X)=\tq(X^2+X^{-2})-S\bigl[\om(\tq^\hf X)-\om(\tq^{-\hf}X)\bigr],
\end{aligned} 
\label{eq:GX-def}
\end{equation}
and $\om(X)$ in \eqref{eq:om-z} is written as
\begin{equation}
\begin{aligned}
 \om(X)=\frac{1}{N}\left\bra\Tr\frac{X-X^{-1}}{X+X^{-1}-A}\right\ket.
\end{aligned} 
\label{eq:omX-def}
\end{equation} 
From this definition, $\om(X)$ satisfies
\begin{equation}
\begin{aligned}
 \om(X^{-1})=-\om(X).
\end{aligned}
\label{eq:om-sym1} 
\end{equation}
Also, from the $\mathbb{Z}_2$ symmetry of our two-matrix model \eqref{eq:VAB-S}
\begin{equation}
\begin{aligned}
 A\to-A,
\end{aligned} 
\end{equation}
$\om(X)$ should satisfy
\begin{equation}
\begin{aligned}
 \om(-X)=\om(X).
\end{aligned} 
\label{eq:om-sym2} 
\end{equation}
From \eqref{eq:om-sym1}, \eqref{eq:om-sym2}, and \eqref{eq:G-eqz},
we find the conditions for $G(X)$
\begin{equation}
\begin{aligned}
 G(X)=G(-X)=G(X^{-1})=G(\tq X).
\end{aligned} 
\label{eq:G-cond}
\end{equation}

For later convenience, we introduce the $q$-difference operator $T_X$
\begin{equation}
\begin{aligned}
 T_Xf(X)=f(\tq^{\hf}X)-f(\tq^{-\hf}X).
\end{aligned} 
\label{eq:q-diff}
\end{equation}
Then $G(X)$ in \eqref{eq:GX-def} is written as
\begin{equation}
\begin{aligned}
 G(X)=\tq (X^2+X^{-2})-ST_X\om(X),
\end{aligned} 
\label{eq:GX-TX}
\end{equation}
and the saddle-point condition \eqref{eq:G-eqz} is written as
\begin{equation}
\begin{aligned}
 T_XG(X)=0.
\end{aligned} 
\label{eq:TX-G}
\end{equation}
\subsection{Solving the saddle-point equation}
As discussed in \cite{Kostov:1999qx,Zakany:2018dio}, 
$G(X)$ obeying the condition \eqref{eq:G-cond}
can be solved by an elliptic function 
by introducing the uniformization coordinate $u$
with the standard double periodicity
\begin{equation}
\begin{aligned}
 u\sim u+1\sim u+\tau.
\end{aligned} 
\end{equation} 
One difference from \cite{Kostov:1999qx,Zakany:2018dio}
is that since $G(X)$ is an even function of
$X$ (see the first equality of \eqref{eq:G-cond}),
the natural variable which is 
uniformized by $u$ is $X^2$, not $X$.
Thus we require
\begin{equation}
\begin{aligned}
 X^2(u+1)=X^2(u)=X^{-2}(-u),\quad X^2(u+\tau)=X^2(u)\tq^2,
\end{aligned} 
\label{eq:X-cond}
\end{equation}
which is satisfied by
\begin{equation}
\begin{aligned}
 X^2(u)=\frac{\vartheta_1(u_0-u)}{\vartheta_1(u_0+u)},
\end{aligned} 
\label{eq:Xu}
\end{equation}
where $\vth_1(u)$ is the Jacobi theta function
\begin{equation}
\begin{aligned}
 \vth_1(u)=-\ri\sum_{n\in\mathbb{Z}}(-1)^n q^{\hf(n+\hf)^2}e^{2\pi\ri u(n+\hf)}
\end{aligned} 
\end{equation}
with $q=e^{2\pi\ri\tau}$. Note that
$\vth_1(u)$ has the properties
\begin{equation}
\begin{aligned}
 \vth_1(u+1)=\vth_1(-u)=-\vth_1(u),\quad
\vth_1(u+\tau)=-e^{-\pi\ri\tau-2\pi\ri u}\vth_1(u).
\end{aligned} 
\label{eq:vth1-sym}
\end{equation}
Note also that $X^2(u)$ in \eqref{eq:Xu} 
has a zero at $u=u_0$ and a pole at $u=-u_0$.
Using \eqref{eq:vth1-sym}, one can show that
the last condition in \eqref{eq:X-cond} is
satisfied by setting $u_0$  as
\begin{equation}
\begin{aligned}
 e^{2\pi\ri u_0}=\tq.
\end{aligned} 
\label{eq:u0-tq}
\end{equation}
One can easily evaluate the value of $X^2(u)$ at several special points
\begin{equation}
\begin{aligned}
 X^2(0)=1,\quad X^2(\pm 1/2)=-1,\quad X^2(\pm\tau/2)=-\tq^{\pm1},\quad
X^2(\pm\tau/2\pm1/2)=-\tq^{\pm1}.
\end{aligned}
\label{eq:X-values} 
\end{equation}

From the definition of $G(X)$ in \eqref{eq:GX-def}, 
$G(X)$ obeys the boundary condition
\begin{equation}
G(X)\sim\left\{
\begin{aligned}
 &\tq X^2+\cO(X^{-2}),&\quad &(X\to\infty),\\
&\tq X^{-2}+\cO(X^2),&\quad &(X\to0).
\end{aligned} \right.
\label{eq:G-asy}
\end{equation}
Also, from \eqref{eq:G-cond} $G(u)=G(X(u))$ should satisfy
\begin{equation}
\begin{aligned}
 G(u+1)=G(-u)=G(u+\tau).
\end{aligned} 
\label{eq:G-peri}
\end{equation} 
One can show that these conditions
\eqref{eq:G-asy} and \eqref{eq:G-peri}
are satisfied by
the following elliptic function 
\begin{equation}
\begin{aligned}
 G(X)&=-\frac{\tq\vth_1(2u_0)}{\vth'_1(0)}
\Biggl[
\del_u\log X^2(u)+\frac{2\vth_1'(2u_0)}{\vth_1(2u_0)}
\Biggr].
\end{aligned} 
\label{eq:GX-result}
\end{equation}
This is our final result of the one-point function,
which implicitly determines $\om(X)$ via the relation  
\eqref{eq:GX-def}.
Note that our result
guarantees that $G(X)$ is an even function of $X$
since both $G(X)$ in \eqref{eq:Xu} and $X^2(u)$ 
in \eqref{eq:Xu} are expanded around $u=\pm u_0$
in the integer powers of $u\mp u_0$.

\subsection{Period integral}
The relation between 
the 't Hooft parameter $S$ and the moduli $\tau$
of the torus is fixed by
the period integral of $G(X)$.
It turns out that the structure of cuts
of $G(X)$ is different for $0<S<1$ and $S>1$.
Here we focus on the $0<S<1$ case.

In this case, there are two segments in the $u$-plane 
depicted by the red and blue lines in the figure below:
\begin{equation}
\begin{aligned}
 \begin{tikzpicture}[scale=1]
\draw[->] (-4,0)--(4,0);
\draw[<-] (0,2.3)--(0,-2.3);
\draw[dashed] (-3,1.5)--(3,1.5);
\draw[dashed] (-3,-1.5)--(3,-1.5);
\draw[dashed] (-3,-1.5)--(-3,1.5);
\draw[dashed] (3,-1.5)--(3,1.5);
\draw[red,thick] (-1.3,1.5)--(-3,1.5);
\draw[red,thick] (3,1.5)--(1.3,1.5);
\draw[blue,thick] (-1.3,-1.5)--(-3,-1.5);
\draw[blue,thick] (3,-1.5)--(1.3,-1.5);
\draw[red,fill=red] (1.3,1.5) circle (.5ex);
\draw[red,fill=red] (-1.3,1.5) circle (.5ex);
\draw[blue,fill=blue] (-1.3,-1.5) circle (.5ex);
\draw[blue,fill=blue] (1.3,-1.5) circle (.5ex);
\draw (0,2.3) node [above]{$\text{Im}(u)$};
\draw (4,0) node [right]{$\text{Re}(u)$};
\draw (-0.1,1.8) node [right]{$\frac{\tau}{2}$};
\draw (1.3,1.5) node [above]{$u_b$};
\draw (-1.3,1.5) node [above]{$\tau-u_b$};
\draw (-1.3,-1.5) node [below]{$-u_b$};
\draw (1.3,-1.5) node [below]{$u_b-\tau$};
\draw (-0.1,-1.8) node [right]{$-\frac{\tau}{2}$};
\draw (2.95,0.3) node [right]{$\hf$};
\draw (-2.95,0.3) node [left]{$-\hf$};
\draw[fill] (0,1) circle (.3ex);
\draw[fill] (0,-1) circle (.3ex);
\draw (0,1) node [right]{$u_0$};
\draw (0,-1) node [right]{$-u_0$};
\end{tikzpicture}
\end{aligned}
\label{eq:cut-fig}
\end{equation}
which correspond to branch cuts of $G(X)$. 
The branch points are located at
\begin{equation}
\begin{aligned}
 u=\pm u_b, \pm(\tau-u_b),
\end{aligned} 
\end{equation}
and $u_b$ is determined by the condition  \cite{Kostov:1999qx}
\begin{equation}
\begin{aligned}
 \del_u\log X^2(u)\Big|_{u=u_b}=\frac{\vth_1'(u_b-u_0)}{\vth_1(u_b-u_0)}
-\frac{\vth_1'(u_b+u_0)}{\vth_1(u_b+u_0)}=0.
\end{aligned} 
\label{eq:ub-def}
\end{equation}

Let us consider the normalization condition
of $G(X)$. Note that $\om(z)$
in \eqref{eq:om-z} should satisfy
\begin{equation}
\begin{aligned}
 \frac{1}{2\pi\ri}\oint_Cdz\om(z)=\frac{1}{2\pi\ri}\oint_C\frac{dX}{X}\om(X)=1,
\end{aligned} 
\end{equation}
where $C$ surrounds the cut of $\om(X)$. This implies that $G(X)$ 
in \eqref{eq:GX-def} satisfies
\begin{equation}
\begin{aligned}
 \Pi_A=\frac{1}{2\pi\ri}\oint_{A}\frac{dX}{X} G(X)=S,
\end{aligned} 
\label{eq:PiA-def}
\end{equation}
where the $A$-cycle surrounds the lower cut 
of $G(X)$ (i.e. the blue line of \eqref{eq:cut-fig}).

We can compute the $A$-period $\Pi_A$ by using the technique in
\cite{Zakany:2018dio} and \cite{Dorey:2002pq}.
As shown in \cite{Zakany:2018dio}, $\Pi_A$  is written as
\begin{equation}
\begin{aligned}
 \Pi_A=\frac{1}{2\pi\ri}\frac{\del}{\del\tau}
\int_{-\hf-\frac{\tau}{2}}^{\hf-\frac{\tau}{2}}duV(\tq^{\hf}X(u))
\end{aligned} 
\label{eq:PA-int}
\end{equation}
where $V(X)$ is the potential for $A$
\begin{equation}
\begin{aligned}
 V(X)=\hf(1-\tq^2)(X+X^{-1})^2.
\end{aligned} 
\end{equation}
The integral \eqref{eq:PA-int} can be evaluated by using the method in \cite{Dorey:2002pq}.
To this end, it is convenient to introduce
\begin{equation}
\begin{aligned}
 f^2(u)=-\frac{\vth_4(u-u_0)}{\vth_4(u+u_0)}
\end{aligned} 
\end{equation}
which is related to $X^2(u)$ in \eqref{eq:Xu} by
\begin{equation}
\begin{aligned}
 f^2(u+\tau/2) =\tq X^2(u).
\end{aligned} 
\label{eq:f-tqX}
\end{equation}
Then $\Pi_A$ in \eqref{eq:PA-int} is written as
\begin{equation}
\begin{aligned}
 \Pi_A&=\frac{1}{2\pi\ri}\frac{\del}{\del\tau}
\int_{-\hf-\frac{\tau}{2}}^{\hf-\frac{\tau}{2}}duV(f(u+\tau/2))\\
&=\frac{1}{2\pi\ri}\frac{\del}{\del\tau}
\int_{-\hf}^{\hf}du'V(f(u')).
\end{aligned} 
\label{eq:f-int}
\end{equation}
From \eqref{eq:f-tqX}, one can show that
\begin{equation}
\begin{aligned}
 X^2(u-\tau/2)-X^2(u+\tau/2)&=(\tq^{-1}-\tq)f^2(u),\\
X^{-2}(u-\tau/2)-X^{-2}(u+\tau/2)&=(\tq-\tq^{-1}) f^{-2}(u),
\end{aligned} 
\end{equation}
and the integral in \eqref{eq:f-int} becomes
\begin{equation}
\begin{aligned}
 \Pi_A&=\frac{1}{2\pi\ri}\frac{\del}{\del\tau}
\int_{-\hf}^{\hf}du \hf (1-\tq^2) \Bigl[f^2(u)+f^{-2}(u)\Bigr]\\
&=\frac{1}{2\pi\ri}\frac{\del}{\del\tau} \hf \tq\int_{-\hf}^{\hf}du \Bigl[
X^2(u-\tau/2)-X^2(u+\tau/2)-X^{-2}(u-\tau/2)+X^{-2}(u+\tau/2)\Bigr].
\end{aligned} 
\end{equation}
As discussed in \cite{Dorey:2002pq}, this integral is evaluated by taking the
residue at $u=\pm u_0$
\begin{equation}
\begin{aligned}
 \Pi_A&=\hf\tq\frac{\del}{\del\tau}\Biggl[\underset{u=-u_0}{\text{Res}}X^2(u)
-\underset{u=u_0}{\text{Res}}X^{-2}(u)\Biggr]\\
&=\tq\frac{\del}{\del\tau}
\frac{\vth_1(2u_0)}{\vth_1'(0)}.
\end{aligned} 
\label{eq:PiA}
\end{equation}
This is our final result of the $A$-period $\Pi_A$.
The condition $\Pi_A=S$ 
\eqref{eq:PiA-def} determines $q=e^{2\pi\ri\tau}$ 
as a function of $\tq$ and $S$.
One can easily show that $\Pi_A$ in \eqref{eq:PiA}
has the following small $q$ expansion
\begin{equation}
\begin{aligned}
 \Pi_A= q (-3 + \tq^{-2} + 3 \tq^2 - \tq^4)+q^2 (-18 + 6\tq^{-2} + 18 \tq^2 - 6 \tq^4) +\cO(q^3).
\end{aligned} 
\label{eq:PiA-small}
\end{equation}
From \eqref{eq:PiA-small} one can see that 
the small $q$ regime corresponds to small $\tq$.
Thus we can compute $q$ as a small $\tq$ expansion with fixed $S$ 
\begin{equation}
\begin{aligned}
 q&=S\tq^2+3S(S-1)^2\tq^4+3S(S-1)^2(9S^2-8S+2)\tq^6+\cO(\tq^8).
\end{aligned} 
\label{eq:q-expand}
\end{equation}
From \eqref{eq:q-expand} and
\eqref{eq:u0-tq}, $\tau$ and $u_0$ are related as
\begin{equation}
\begin{aligned}
 \tau\approx 2u_0\qquad(\tq\to0).
\end{aligned} 
\end{equation}
Note also that $\tau$ and $u_0$ are pure imaginary when $0<\tq<1,0<S<1$.

We can also compute $u_b$ in the small $\tq$ expansion
by solving the equation for the branch-point \eqref{eq:ub-def}
\begin{equation}
\begin{aligned}
 e^{2\pi\ri u_b}&=\rt{S}(\rt{S}-\rt{S-1})\tq\left[1
+\hf(S-1)^{3/2}(3\rt{S-1}-5\rt{S})\tq^2+\cO(\tq^4)\right].
\end{aligned} 
\label{eq:ub-sol}
\end{equation}
If we set $S=\cos^2\phi$, 
$u_b$ in \eqref{eq:ub-sol} is expanded as
\begin{equation}
\begin{aligned}
u_b&=\frac{\tau}{2}+\text{Re}(u_b),\\
\text{Re}(u_b)&=\frac{1}{4\pi}\bigl[2\phi-5\tq^2\cos\phi\sin^3\phi+\cO(\tq^4)\bigr].
\end{aligned} 
\label{eq:ub}
\end{equation}
It is also convenient to define
\begin{equation}
\begin{aligned}
 X_b=X(u_b)\tq^{-\hf},\quad
\tilde{S}=\left(\frac{X_b+X_b^{-1}}{2}\right)^2.
\end{aligned} 
\end{equation}
Then we find that $\tilde{S}$ is expanded as
\begin{equation}
\begin{aligned}
\tilde{S}
=S+S(S-1)^2\tq^2+S(S-1)^2(1-6S+8S^2)\tq^4+
\cO(\tq^6).
\end{aligned} 
\label{eq:def-tilS}
\end{equation}

\subsection{Moment of $\om(X)$}
Let us consider the small $X$ expansion of $\om(X)$
in \eqref{eq:omX-def}
\begin{equation}
\begin{aligned}
 \om(X)=-\sum_{n=0}^\infty\mu_n X^n.
\end{aligned}
\label{eq:om-smallX} 
\end{equation}
From the generating function of the Chebyshev polynomials $T_n(t)$ 
\begin{equation}
\frac{1-X^2}{1-2tX+X^2}=1+\sum_{n=1}^\infty 2T_n(t)X^n,
\label{eq:Tn-gen}
\end{equation}
one can see that the coefficient $\mu_n$ of this expansion \eqref{eq:om-smallX} 
is the expectation value of the moment
\begin{equation}
\begin{aligned}
 \mu_n=\frac{1}{N}\Bigl\bra\Tr \bigl(2T_n(A/2)\bigr)
\Bigr\ket,\quad (n\geq1).
\end{aligned} 
\label{eq:def-mu}
\end{equation}
Plugging \eqref{eq:om-smallX} into the definition of $G(X)$ in \eqref{eq:GX-def}, 
$G(X)$ is expanded as
\begin{equation}
\begin{aligned}
 G(X)=\tq(X^2+X^{-2})+S\sum_{n=1}^\infty \mu_n
(\tq^{\frac{n}{2}}-\tq^{-\frac{n}{2}})X^n.
\end{aligned} 
\end{equation}
We can easily extract the moment $\mu_n$
from the expansion of our solution of $G(X)$ \eqref{eq:GX-result}
around $u=u_0$.
Our solution \eqref{eq:GX-result}
guarantees that the odd moments vanish
\begin{equation}
\begin{aligned}
 \mu_{2k+1}=0,\quad (k\in\mathbb{Z}_{\geq0}),
\end{aligned} 
\end{equation}
since $G(X)$ in \eqref{eq:GX-result} is an even function of $X$ by construction.
Thus $G(X)$ is expanded as
\begin{equation}
\begin{aligned}
 G(X)=
\tq(X^2+X^{-2})+S\sum_{n=1}^\infty \mu_{2n}
(\tq^{n}-\tq^{-n})X^{2n}.
\end{aligned} 
\end{equation}
For instance, $\mu_2$ and $\mu_4$ are obtained from \eqref{eq:GX-result} as
\begin{equation}
\begin{aligned}
 \mu_2&=\frac{\tq}{S(\tq-\tq^{-1})}
\left[\frac{\vth_1'''(0)\vth_1(2u_0)^2}{2\vth_1'(0)^3}+\frac{\vth_1'(2u_0)^2}{\vth_1'(0)^2}-\frac{3\vth_1(2u_0)\vth_1''(2u_0)}{2\vth_1'(0)^2}-1\right],\\
\mu_4&=\frac{2\tq \vth_1(2u_0)^2}{3S(\tq^2-\tq^{-2})}\left[
\frac{\vth_1'''(2u_0)}{\vth_1'(0)^3}-
\frac{\vth_1'''(0)\vth_1'(2u_0)}{\vth_1'(0)^4}\right].
\end{aligned} 
\end{equation}
Using the small $\tq$ expansion of $q$ in \eqref{eq:q-expand},
we can compute the small $\tq$ expansion of the moments
$\mu_{2n}$
\begin{equation}
\begin{aligned}
 \mu_2&=S-2+S(S-1)^2\tq^2+S(S-1)^2(6S^2-6S+1)\tq^4+\cO(\tq^6),\\
\mu_4&=2(S-1)^2+4(S-1)^3S\tq^2+2S(S-1)^2(14S^3-26S^2+15S-2)\tq^4+\cO(\tq^6),\\
\mu_6&=
(-1+S)^2 (-2+5 S)+3 (-1+S)^3 S (-3+5 S) \tq^2\\
&\quad +9 (-1+S)^3 S \left(-1+9 S-19 S^2+13 S^3\right) \tq^4+\cO(\tq^6).
\end{aligned} 
\label{eq:mu-result}
\end{equation}
One can check that the $\cO(\tq^0)$ term of $\mu_{2n}$ is equal to
\begin{equation}
\begin{aligned}
 \lim_{\tq\to0}\mu_{2n}=\int_{-2\rt{S}}^{2\rt{S}}d\la\rho(\la)
2T_{2n}(\la/2),
\end{aligned} 
\label{eq:lim-mu}
\end{equation}
where $\rho(\la)$ is the eigenvalue density of Gaussian one-matrix model
with the potential $V(A)=\frac{A^2}{2S}$
\begin{equation}
\begin{aligned}
 \rho(\la)=\frac{\rt{4S-\la^2}}{2\pi S}.
\end{aligned} 
\end{equation}
This is expected since in the limit $\tq\to0$
the two-matrix model in \eqref{eq:VAB-S} reduces
to two decoupled Gaussian one-matrix models
for $A$ and $B$
\begin{equation}
\begin{aligned}
 \lim_{\tq\to0}\Tr V(A,B)=\Tr\left(\frac{A^2}{2S}+\frac{B^2}{2}\right).
\end{aligned} 
\label{eq:tq0-lim}
\end{equation}

It turns out that the higher order correction in \eqref{eq:mu-result}
can be expressed as a deformation of the eigenvalue density
\begin{equation}
\begin{aligned}
 \rho(\la)=\frac{\rt{4\tilde{S}-\la^2}}{2\pi\tilde{S}}
\frac{1+c\la^2+
\cO(\la^4)}{1+c\tilde{S}+\cdots},\end{aligned} 
\end{equation}
where $\tilde{S}$ is defined in \eqref{eq:def-tilS} and $c$ is 
given by
\begin{equation}
\begin{aligned}
 c=-2S(S-1)^2\tq^4+\cO(\tq^6).
\end{aligned} 
\end{equation}

\subsection{One-point function at $S=1$}
We observe from \eqref{eq:mu-result}
that the $S\to1$ limit of the moments is given by
\begin{equation}
\begin{aligned}
 \lim_{S\to1}\mu_n=-\cob_{n,2},
\end{aligned} 
\end{equation}
which is independent of $\tq$.
This is consistent with the result of \cite{Jafferis:2022wez}
that the planar one-point function of the two-matrix
model at $S=1$ remains the same as that of the one-matrix model
with the potential $V_0(A)$ in \eqref{eq:V0-Vct}, which reduces to the Gaussian one-matrix model when $q_A=0$
\begin{equation}
\begin{aligned}
 \lim_{q_A\to0}V_0(A)=T_2(A/2)=\frac{A^2}{2}-1.
\end{aligned} 
\label{eq:V0-lim}
\end{equation}

We can explicitly show that the saddle-point condition \eqref{eq:G-cond}
for $S=1$ is solved by the resolvent of Gaussian one-matrix model with the eigenvalue
density
\begin{equation}
\begin{aligned}
 \rho(\la)=\frac{\rt{4-\la^2}}{2\pi}.
\end{aligned} 
\end{equation}
The resolvent \eqref{eq:resolvent} for this eigenvalue density is given by
\begin{equation}
\begin{aligned}
R(\la)=\int_{-2}^2d\la'\rho(\la')\frac{1}{\la-\la'}=\hf\bigl(\la\pm \rt{\la^2-4}\bigr)
=X^{\pm1},
\end{aligned} 
\label{eq:G-resolvent}
\end{equation}
where we used the relation between $\la$ and $X$ in \eqref{eq:la-X}.
$X$ and $X^{-1}$ in \eqref{eq:G-resolvent} correspond to
different branches of the resolvent.
From \eqref{eq:G-resolvent} and the definition of $\om(X)$ in \eqref{eq:omX-def}, 
we find the 
expansion of $\om(X)$ around $X=0$ and $X=\infty$
\begin{equation}
\om(X)=\left\{
\begin{aligned}
 \om_0(X)&=X(X-X^{-1})=X^2-1,&\quad &(X\to0),\\
\om_\infty(X)&=X^{-1}(X-X^{-1})=1-X^{-2},&\quad &(X\to\infty).
\end{aligned} \right.
\label{eq:om-S1}
\end{equation}
Then we find
\begin{equation}
\begin{aligned}
 T_X\om(X)&=\om_0(\tq^{\hf}X)-\om_\infty(\tq^{-\hf}X)\\
&=\tq(X^2+X^{-2})-2,
\end{aligned} 
\end{equation}
where we assumed
\begin{equation}
\begin{aligned}
 \tq^\hf\ll \min\bigl\{|X|,|X^{-1}|\bigr\}.
\end{aligned} 
\end{equation}
Finally, we find that $G(X)$ in \eqref{eq:GX-TX}
for $S=1$ is constant
\begin{equation}
\begin{aligned}
 G(X)=\tq(X^2+X^{-2})-T_X\om(X)=2,
\end{aligned} 
\end{equation}
which trivially satisfies the saddle-point condition \eqref{eq:G-cond}.

We should stress that the two-matrix model with the potential
\eqref{eq:VAB} itself
is not equal to the Gaussian matrix model when $\tq\ne0$. 
We are only claiming 
that the planar 
resolvent $\om(X)$
in \eqref{eq:om-S1} for $S=1$ 
is equal to that of the Gaussian one-matrix model \eqref{eq:V0-lim},
which is expected from the construction of the counter term \eqref{eq:Vct}.

\subsection{Structure of $u$-plane}\label{sec:one-S1}
From the result of $\Pi_A$ in \eqref{eq:PiA},
one can show that the condition $\Pi_A=S$ for $S=1$
is exactly solved by 
\begin{equation}
\begin{aligned}
 q=\tq^2.
\end{aligned} 
\end{equation}
Namely, $\tau=2u_0$ for $S=1$.
This is a singular limit since $X(u)$ in \eqref{eq:Xu} becomes
an elementary function $X(u)=e^{\pi\ri u}$ when $q=\tq^2$.
There are no zero and pole for $X(u)=e^{\pi\ri u}$. When we take a limit $S\to1$
from the $S<1$ side, 
what is happening is that
the zero at $u=u_0$ and the pole at $u=\tau-u_0$ collide as $u_0\to\tau/2$
and they pair-annihilate and disappear.
Then, the structure of $u$-plane for the $S=1$ case is depicted as
\begin{equation}
\begin{aligned}
 \begin{tikzpicture}[scale=1]
\draw[->] (-4,0)--(4,0);
\draw[<-] (0,2.3)--(0,-2.3);
\draw[dashed] (-3,1.5)--(3,1.5);
\draw[dashed] (-3,-1.5)--(3,-1.5);
\draw[dashed] (-3,-1.5)--(-3,1.5);
\draw[dashed] (3,-1.5)--(3,1.5);
\draw[red,thick] (-3,1.5)--(3,1.5);
\draw[blue,thick] (-3,-1.5)--(3,-1.5);
\draw (0,2.3) node [above]{$\text{Im}(u)$};
\draw (4,0) node [right]{$\text{Re}(u)$};
\draw (-0.1,1.8) node [right]{$\frac{\tau}{2}$};
\draw (3,1.5) node [above]{$\frac{\tau}{2}+\hf$};
\draw (-3,1.5) node [above]{$\frac{\tau}{2}-\hf$};
\draw (-3,-1.5) node [below]{$-\frac{\tau}{2}-\hf$};
\draw (3,-1.5) node [below]{$-\frac{\tau}{2}+\hf$};
\draw (-0.1,-1.8) node [right]{$-\frac{\tau}{2}$};
\draw (2.95,0.3) node [right]{$\hf$};
\draw (-2.95,0.3) node [left]{$-\hf$};
\end{tikzpicture}
\end{aligned}
\label{eq:cut-S1}
\end{equation}

\begin{figure}[t]
\centering
\includegraphics
[width=0.45\linewidth]{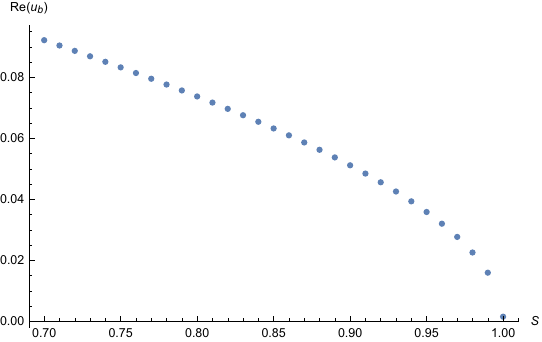}
  \caption{
Plot 
of $\text{Re}(u_b)$ as a function of $S$ for $\tq=0.3$.
}
  \label{fig:ub}
\end{figure}

In Figure~\ref{fig:ub}, we show the plot of $\text{Re}(u_b)$
computed numerically for $\tq=0.3$. We can see that
$\text{Re}(u_b)$ vanishes at $S=1$,
which implies that the gap around $u=\tau/2$
in \eqref{eq:cut-fig} closes as $S\to1$.
This is reminiscent of the Gross-Witten transition of 
unitary matrix model \cite{Gross:1980he}.\footnote{We do not understand whether the free energy of our 
two-matrix model is singular at $S=1$ or not.
It might be the case that the apparent singularity at $S=1$ is an artifact of our
parametrization of the eigenvalue $\la=X(u)+X(u)^{-1}$.}

\begin{figure}[t]
\centering
\includegraphics
[width=0.45\linewidth]{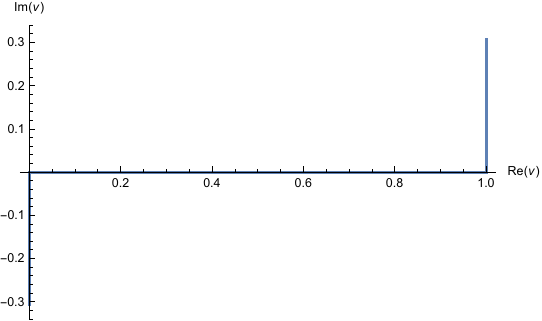}
  \caption{
Plot 
of $v$ in \eqref{eq:v-def} for $\la\in[-3,3]$.
}
  \label{fig:logX}
\end{figure}

The $u$-plane diagram in \eqref{eq:cut-fig} is valid for $0<S<1$.
We do not have a complete understanding of the solution of $G(X)$ for $S>1$.
We can speculate the structure of $u$-plane for $S>1$
by assuming that $X(u)\approx e^{\pi \ri u}$ near $S=1$.
Then it is natural to define $v$ as
\begin{equation}
\begin{aligned}
 v=\frac{1}{\pi\ri}\log X=\frac{1}{\pi\ri}\log\frac{\la+\rt{\la^2-4}}{2},
\end{aligned} 
\label{eq:v-def}
\end{equation}
which is an analogue of the variable $u$. 
When $S>1$, we expect that the end-point of the cut on the $\la$-plane 
is larger than $2$.
In Figure~\ref{fig:logX}, we show the plot of $v$ in \eqref{eq:v-def}
for $\la\in[-3,3]$. 
From this behavior, we conjecture that the structure of $u$-plane for $S>1$
looks like this:
\begin{equation}
\begin{aligned}
 \begin{tikzpicture}[scale=1]
\draw[->] (-4,0)--(4,0);
\draw[<-] (0,2.3)--(0,-2.3);
\draw[dashed] (-3,1.5)--(3,1.5);
\draw[dashed] (-3,-1.5)--(3,-1.5);
\draw[dashed] (-3,-1.5)--(-3,1.5);
\draw[dashed] (3,-1.5)--(3,1.5);
\draw[red,thick] (-3,1.5)--(3,1.5);
\draw[red,thick] (0,1)--(0,2);
\draw[blue,thick] (-3,-1.5)--(3,-1.5);
\draw[blue,thick] (0,-1)--(0,-2);
\draw[red,fill=red] (0,2) circle (.5ex);
\draw[red,fill=red] (0,1) circle (.5ex);
\draw[blue,fill=blue] (0,-2) circle (.5ex);
\draw[blue,fill=blue] (0,-1) circle (.5ex);
\draw (0,2) node [right]{$u_b$};
\draw (0,1) node [left]{$\tau-u_b$};
\draw (0,-2) node [right]{$-u_b$};
\draw (0,-1) node [left]{$u_b-\tau$};
\draw (0,2.3) node [above]{$\text{Im}(u)$};
\draw (4,0) node [right]{$\text{Re}(u)$};
\draw (2.95,0.3) node [right]{$\hf$};
\draw (-2.95,0.3) node [left]{$-\hf$};
\draw[fill] (0,0.5) circle (.3ex);
\draw[fill] (0,-0.5) circle (.3ex);
\draw (0,0.5) node [right]{$\tau-u_0$};
\draw (0,-0.5) node [right]{$u_0-\tau$};
\end{tikzpicture}
\end{aligned}
\label{eq:cut-S>1}
\end{equation}
It would be interesting to find the analytic solution of $G(X)$ for $S>1$.
We leave this as an interesting future problem.

\section{Two-point function of resolvents}\label{sec:two}

In this section, we study the two-point function of resolvents in our two-matrix
model  \eqref{eq:VAB-S}.
In a similar manner as 
the one-point function $\omega(X)$ in \eqref{eq:omX-def}, we define the 
connected two-point function $\omega(X,Y)$ by 
\begin{equation}
\begin{aligned}
 \omega(X,Y)&=\left\langle \Tr\frac{X-X^{-1}}{X+X^{-1}-A}\Tr\frac{Y-Y^{-1}}{Y+Y^{-1}-A} \right\rangle_{\rm conn}\\
&= \left\langle \sum_{a=1}^N\frac{X-X^{-1}}{X+X^{-1}-\la_a}
\sum_{b=1}^N\frac{Y-Y^{-1}}{Y+Y^{-1}-\la_b} \right\rangle_{\rm conn}.
\end{aligned} 
\label{eq:om2pt-def}
\end{equation}
Our definition of $\om(X,Y)$ in \eqref{eq:om2pt-def}
implies the following symmetry properties: 
\begin{equation}
\omega(X,Y)=\omega(Y,X)=-\omega(X^{-1},Y)=\omega(-X,-Y). 
   \label{2pt symmetry}
\end{equation}
Note that $\omega(-X,Y)$ is not equal to $\omega(X,Y)$ in general.
Naively, we expect that the structure of cuts of $\omega(X,Y)$
is inherited from that of the one-point function $\omega(X)$. 
However, this is not the case due to the fact that 
$\omega(-X,Y)\ne \om(X,Y)$. We find that
only the even part of $\omega(X,Y)$ can be expressed in a simple way
for general $S<1$.
Interestingly, we find that $S=1$ is special for the two-point function as well.
It turns out that when $S=1$ the entire $\omega(X,Y)$ 
can be constructed in terms of the Bergman kernel on a torus.

\subsection{Equation for $\omega(X,Y)$}\label{sec:2pt-eq}
In this subsection,
we derive an equation satisfied by $\omega(X,Y)$. 
A similar equation is discussed, for example, in \cite{Zakany:2018dio}. 

For this purpose, we need to extend the saddle-point equation \eqref{eq:saddle} 
to the matrix model with a generic potential for $A$ 
\begin{equation}
V(X;t)=\sum_{n\in \mathbb{Z}}t_nX^n. 
\end{equation}
Here the 't Hooft parameter $S$ is absorbed into the definition of
the parameters $t_n$ in the potential $V(X;t)$.
The one-point function of this model is 
given by
\begin{equation}
\omega_t(X)=\frac1N\left\langle \sum_{a=1}^N\frac{X-X^{-1}}{X+X^{-1}-X_a-X_a^{-1}} \right\rangle_t, 
\label{eq:om-t}
\end{equation}
where $X_a$ is related to the eigenvalue $\la_a$ of $A$ by 
\begin{equation}
\lambda_a\ =\ X_a+X_a^{-1}, 
\end{equation}
and the average $\langle\cdots\rangle_t$ is taken with respect to $V(X;t)$
\begin{equation}
\begin{aligned}
 \langle\cdots\rangle_t=\frac{1}{\cZ_t}\int\prod_{a=1}^N dz_a e^{-NV(e^{z_a};t)}(\cdots)\prod_{a<b}
\frac{(\cosh z_a-\cosh z_b)^2}{\bigl(\cosh(z_a+\lap)-\cosh z_b\bigr)
\bigl(\cosh(z_a-\lap)-\cosh z_b\bigr)},
\end{aligned} 
\end{equation} 
where $\cZ_t$ is determined by the condition $\bra 1\ket_t=1$.
The saddle-point equation for $\omega_t(X)$ turns out to be 
\begin{equation}
\sum_{n\in\mathbb{Z}}nt_nX^n=2\omega_t(X)-\omega_t(\tilde{q}X)-\omega_t(\tilde{q}^{-1}X). 
   \label{saddle_t}
\end{equation}

From \eqref{eq:Tn-gen}, one can show the following identity 
\begin{equation}
\frac{Y-Y^{-1}}{Y+Y^{-1}-X-X^{-1}}=-\sum_{n\in\mathbb{Z}}X^nY^{|n|}. 
\label{eq:res-XY}
\end{equation}
Introducing 
the so-called
loop-insertion operator $D(Y)$ by
\begin{equation}
\begin{aligned}
 D(Y)=\sum_{n\in\mathbb{Z}}Y^{|n|}\frac\partial{\partial t_n},
\end{aligned} 
\label{eq:def-DY}
\end{equation}
\eqref{eq:res-XY} is written as
\begin{equation}
\begin{aligned}
 \frac{Y-Y^{-1}}{Y+Y^{-1}-X-X^{-1}}=-D(Y)V(X;t).
\end{aligned} 
\end{equation}
Then, acting $D(Y)$ on $\om_t(X)$ in \eqref{eq:om-t} we find  
\begin{equation}
D(Y)\omega_t(X)=\omega_t(X,Y), 
   \label{1 to 2}
\end{equation}
where $\omega_t(X,Y)$ is the two-point function defined with respect to $V(X;t)$.
Our model of interest 
corresponds to $t_n=t_n^*$ such that 
\begin{equation}
V(X;t^*)\ =\ \frac{1}{2S}(1-\tq^2)(X+X^{-1})^2,
\end{equation} 
and we denote $\om(X)=\om_{t^*}(X)$ and $\om(X,Y)=\om_{t^*}(X,Y)$.
Using the relation \eqref{1 to 2},
one can obtain an equation for $\omega(X,Y)$
by acting the loop-insertion operator $D(Y)$ on both sides of \eqref{saddle_t}
and setting $t_n=t_n^*$.
Acting $D(Y)$ on the left-hand side of \eqref{saddle_t}, we find
\begin{equation}
D(Y)\sum_{n\in\mathbb{Z}}nt_nX^n=
-K(X,Y)+K(X,Y^{-1}), 
\end{equation}
where 
\begin{equation}
K(X,Y)=\frac{XY}{(X-Y)^2}, 
\label{eq:KXY}
\end{equation}
and the right-hand side of \eqref{saddle_t} becomes
\begin{equation}
\begin{aligned}
 D(Y)\bigl[2\omega_t(X)-\omega_t(\tilde{q}X)-\omega_t(\tilde{q}^{-1}X)\bigr]
=2\om_t(X,Y)-\om_t(\tq X,Y)-\om_t(\tq^{-1}X,Y).
\end{aligned} 
\end{equation}
Finally we find the equation for $\omega(X,Y)$
\begin{equation}
\begin{aligned}
 -K(X,Y)+K(X,Y^{-1})&=2\omega(X,Y)-\omega(\tilde{q}X,Y)-\omega(\tilde{q}^{-1}X,Y)\\
&=-T_X^2\om(X,Y), 
   \label{2pt saddle}
\end{aligned} 
\end{equation}
where we define $T_X$ and $T_Y$ 
as $q$-difference operators
generalizing \eqref{eq:q-diff} to the two-variable case
\begin{equation}
\begin{aligned}
 T_Xf(X,Y) &= f(\tilde{q}^{\frac12}X,Y)-f(\tilde{q}^{-\frac12}X,Y), \\
T_Yf(X,Y) &= f(X,\tilde{q}^{\frac12}Y)-f(X,\tilde{q}^{-\frac12}Y). 
\end{aligned} 
\end{equation}
Note that the left-hand side 
of \eqref{2pt saddle} 
is compatible with the symmetry \eqref{2pt symmetry} of $\omega(X,Y)$. 

As in the case of one-point function $\omega(X)$, this equation 
\eqref{2pt saddle} can be simplified by introducing another function $G(X,Y)$ defined by 
\begin{equation}
G(X,Y)=T_XT_Y\omega(X,Y)+K(X,Y)+K(X,Y^{-1}). 
   \label{def G}
\end{equation}
From the symmetries of
$\om(X,Y)$ in \eqref{2pt symmetry}, it follows that
$G(X,Y)$ has the symmetries
\begin{equation}
G(X,Y)=G(Y,X)=G(X^{-1},Y)=G(-X,-Y). 
\end{equation}

Acting $T_Y$ on both sides of \eqref{2pt saddle}
and using the relation
\begin{equation}
T_YK(X,Y)=-T_XK(X,Y), \hspace{5mm} T_YK(X,Y^{-1})=T_XK(X,Y^{-1}),
\end{equation}
we find the equation for $G(X,Y)$
\begin{equation}
T_XG(X,Y)=0. 
   \label{G saddle}
\end{equation}
For a fixed $Y$, this equation is the same as \eqref{eq:TX-G} for $G(X)$. 
Therefore, this equation can also be converted to the periodicity condition by employing the conformal transformation $X=X(u)$ and $Y=X(v)$ defined in \eqref{eq:Xu}. 
The function $G(u,v)=G(X(u),X(v))$ then satisfies 
\begin{equation}
G(u,v)=G(u+1,v)=G(u+\tau,v). 
\end{equation}
The same periodicity condition is satisfied for $v$. 

However, it turns out that $G(u,v)$ is not an elliptic function of $u$ nor $v$
by the following reason:
Note that $G(X,Y)$ is not invariant under $X\to-X$ while $Y$ is kept fixed. 
From the definition  of $X(u)$ in \eqref{eq:Xu}, we have
\begin{equation}
\begin{aligned}
 X(u)=\rt{\frac{\vth_1(u_0-u)}{\vth_1(u_0+u)}}.
\end{aligned} 
\end{equation}
One can see that $X(u)$ has square-root branch points at $u=\pm u_0$.
Thus, the transformation $X\to-X$ 
is realized by moving $u$ along a closed path encircling $u=u_0$ 
and coming back to the original point on the $u$-plane. 
The non-invariance of $G(u,v)$ under this transformation then suggests that 
$G(u,v)$ has an extra branch cut on the $u$-plane, in addition to the ones 
indicated by the red and blue lines in \eqref{eq:cut-fig}.
Therefore, $G(u,v)$ would be rather a function on a Riemann surface of genus-two 
in general. 

It is more convenient to consider the even and the odd parts of $G(X,Y)$ defined by 
\begin{equation}
G_\pm(X,Y)=\frac12\Bigl[ G(X,Y)\pm G(-X,Y) \Bigr]. 
\end{equation}
Then, the even part $G_+(X,Y)$ satisfies $G_+(-X,Y)=G_+(X,Y)$
and hence it is an elliptic function. 
All the complications are contained in the odd part $G_-(X,Y)$ which we will not discuss in this paper. 

When $S=1$, the above problem is curcumvented by the fact that
$X(u)=e^{\pi\ri u}$ is an entire function on the $u$-plane.
As a consequence, the whole $G(X,Y)$ becomes an elliptic function when $S=1$.

\subsection{Even part of $G(X,Y)$ for $S<1$}

In this subsection, we determine the explicit form of the even part 
$G_+(X,Y)$
for $S<1$. 
From the definition \eqref{def G} of $G(X,Y)$ we find
\begin{equation}
G_+(X,Y)=\frac{2X^2Y^2}{(X^2-Y^2)^2}+\frac{2X^2Y^{-2}}{(X^2-Y^{-2})^2}+T_XT_Y\omega_+(X,Y), 
\end{equation}
where $\omega_+(X,Y)$ is the even part of $\omega(X,Y)$. 
This shows that $G_+(X,Y)$ has double poles at $X=Y,Y^{-1}$ whose structures are given by 
\begin{equation}
\lim_{X\to Y^{\pm1}}G_+(X,Y)\sim\frac{2X^2Y^{\pm2}}{(X^2-Y^{\pm2})^2}. 
\label{eq:G+-pole}
\end{equation}
These conditions can be solved by
an elliptic function, known as the Bergman kernel on a torus \cite{Eynard:2008we,Zakany:2018dio}, given by 
\begin{equation}
B(X,Y)=X\partial_XY\partial_Y\log\vartheta_1(u_X-u_Y), 
\end{equation}
where $u_X=u(X)$ is the inverse function of $X=X(u)$
in \eqref{eq:Xu}. 
If $u_X$ approaches $u_Y$, $B(X,Y)$ behaves as 
\begin{equation}
B(X,Y)\sim X\partial_XY\partial_Y\log(X^2-Y^2)=4K(X^2,Y^2). 
\label{eq:B-asy}
\end{equation}
From \eqref{eq:G+-pole}
and \eqref{eq:B-asy}, we find that $G_+(X,Y)$ is given by
\begin{equation}
\begin{aligned}
 G_+(X,Y)&=\hf B(X,Y)+\hf B(X,Y^{-1})\\
&=\frac12X\partial_XY\partial_Y\log\frac{\vartheta_1(u_X-u_Y)}{\vartheta_1(u_X+u_Y)},
\end{aligned} 
\label{eq:G+-bergman}
\end{equation}
where we used $X^{2}(-u)=X^{-2}(u)$.
This is also written as
\begin{equation}
\begin{aligned}
G_+(X,Y)
&=& 2\frac{\partial_{u_X}\partial_{u_Y}[\log\vartheta_1(u_X-u_Y)-\log\vartheta_1(u_X+u_Y)]}{\partial_{u_X}\log X^2\partial_{u_Y}\log Y^2},
   \label{G_+}
 \end{aligned} 
\end{equation}
which makes manifest that $G_+(X,Y)$ is
an elliptic function for both variables $u_X$ and $u_Y$.

Based on the relation \eqref{1 to 2}, we expect that $G_+(u,v)$ has a simple pole at the branch point $u=u_b$. 
This can be seen as follows. 
The one-point function $\omega_t(X)$ behaves as 
\begin{equation}
\omega_t(X)\sim\omega_t(X_b(t))+c\sqrt{X-X_b(t)}
\end{equation}
near a branch point $X=X_b(t)$ with some constant $c$ and $X_b(t^*)=X(u_b)$. 
Since $\omega(X,Y)$ is obtained from the $t_n$-derivatives of $\omega_t(X)$
as shown in \eqref{1 to 2}, it should behave near the branch point as 
\begin{equation}
\omega(X,Y)\sim\frac {c'}{\sqrt{X-X(u_b)}}. 
\end{equation}
This corresponds to a simple pole for $u_X$ at $u_X=u_b$,
since $X'(u_b)=0$ by definition of $u_b$ in \eqref{eq:ub-def}
and hence $X-X_b$ is of order $(u_X-u_b)^2$.
Since the residue of the simple pole is unknown, the analytic structure of $G_+(X,Y)$ alone cannot determine its functional form completely. 
This ambiguity is fixed by requiring
that the periods of $G_+(X,Y)$ around the branch cuts vanish, 
which is expected from the definition \eqref{def G}.
One can check that the $A$-period of our solution of $G_+(X,Y)$ 
in \eqref{G_+} indeed vanishes.\footnote{In fact, the Bergman kernel 
is defined to have a vanishing 
integral around the A-cycle \cite{Eynard:2008we}.}

\subsection{$G(X,Y)$ at $S=1$}

Let us return to $G(X,Y)$ itself. 
When $S=1$, $G(X,Y)$ becomes an elliptic function
as we discussed at the end of subsection \ref{sec:2pt-eq}.

We find that $G(X,Y)$ at $S=1$ is given by the following infinite sum: 
\begin{equation}
G(X,Y)=\sum_{m\in\mathbb{Z}}\Bigl[ K(\tilde{q}^mX,Y)+K(\tilde{q}^mX,Y^{-1}) \Bigr].
   \label{infinite sum}
\end{equation}
This obviously satisfies the equation \eqref{G saddle} and this has the correct pole structures at $X=Y,Y^{-1}$. We can also check that
the even part of \eqref{infinite sum}
agrees with the $S\to1$ limit of $G_+(X,Y)$ in \eqref{eq:G+-bergman},
as we will see below.

From \eqref{infinite sum} and \eqref{def G},  $T_XT_Y\omega(X,Y)$
is given by
\begin{equation}
\begin{aligned}
 T_XT_Y\omega(X,Y)&=\sum_{m\in\mathbb{Z},m\ne0}\Bigl[K(\tilde{q}^mX,Y)+K(\tilde{q}^mX,Y^{-1})\Bigr]\\
&=\sum_{n=1}^\infty \frac{n\tilde{q}^n}{1-\tilde{q}^n}(X^n+X^{-n})(Y^n+Y^{-n}), 
\end{aligned} 
   \label{TTomega}
\end{equation}
where we assumed $\tq$ is small and we used the expansion of $K(X,Y)$
\begin{equation}
K(X,Y)=\left\{
\begin{aligned}
&\sum_{n=1}^\infty nX^nY^{-n},\quad &(|X|<|Y|),\\
& \sum_{n=1}^\infty nX^{-n}Y^{n},\quad &(|X|>|Y|).
\end{aligned} \right.
\end{equation}
From \eqref{TTomega} we can compute the 
moments of two-point function as follows.
Using \eqref{eq:Tn-gen},
$\omega(X,Y)$ can be expanded for small $X$ and $Y$ as 
\begin{equation}
\omega(X,Y)=\sum_{n,m=1}^\infty c_{n,m}X^nY^m, 
   \label{expand 2pt}
\end{equation}
where 
\begin{equation}
c_{n,m}=\left\langle {\rm Tr}\left( 2T_n\left( \frac A2 \right) \right){\rm Tr}\left( 2T_m\left( \frac A2 \right) \right) \right\rangle_{\rm conn}. 
\label{eq:cnm}
\end{equation}
The expression \eqref{TTomega} implies that $c_{n,m}$ vanishes for $n\ne m$
\begin{equation}
\begin{aligned}
 c_{n,m}=c_n\delta_{n,m}, 
\end{aligned} 
\label{eq:cnm-cn}
\end{equation}
and \eqref{expand 2pt} becomes
\begin{equation}
\begin{aligned}
 \omega(X,Y)=\sum_{n=1}^\infty c_{n}X^nY^n. 
\end{aligned} 
\label{eq:om-cn}
\end{equation}
Plugging 
\eqref{eq:om-cn}
into the explicit form of $T_XT_Y\omega(X,Y)$
\begin{equation}
T_XT_Y\omega(X,Y)=\omega(\tilde{q}^{\frac12}X,\tilde{q}^{\frac12}Y)+\omega(\tilde{q}^{-\frac12}X,\tilde{q}^{-\frac12}Y)-\omega(\tilde{q}^{\frac12}X,\tilde{q}^{-\frac12}Y)-\omega(\tilde{q}^{-\frac12}X,\tilde{q}^{\frac12}Y), 
\label{eq:TXTY-S1}
\end{equation}
and using the symmetry \eqref{2pt symmetry}, we obtain 
\begin{equation}
T_XT_Y\omega(X,Y)=\sum_{n=1}^\infty c_n\tilde{q}^n(X^n+X^{-n})(Y^n+Y^{-n}),
\end{equation}
where we assumed that $X,Y$ satisfy 
\begin{equation}
\tilde{q}^\frac12\ll \min\{|X|,|X^{-1}|,|Y|,|Y^{-1}|\}. 
\end{equation}
Comparing \eqref{eq:TXTY-S1} with \eqref{TTomega}, we find
\begin{equation}
c_n=\frac n{1-\tilde{q}^n}. 
   \label{c_n}
\end{equation}
Remarkably, this reproduces the result of \cite{Jafferis:2022wez}
for the moments of two-point function at $S=1$
(see eq.(9.29) in \cite{Jafferis:2022wez}). 
We should stress that our derivation of $c_n$ in \eqref{c_n} is 
different from \cite{Jafferis:2022wez}.
In \cite{Jafferis:2022wez}, $c_n$ was obtained by summing over the effect
of matter loops as a geometric series
\footnote{The factor $\frac{\tq^n}{1-q_A^n}$ of matter loop
also appeared in the half-wormhole amplitude \cite{Okuyama:2023byh}
with the identification
$\tq=a, n=\mathfrak{b}$.}
\begin{equation}
\begin{aligned}
 c_n=n\sum_{m=0}^\infty\left(\frac{\tq^n}{1-q_A^n}\right)^m
=n\frac{1-q_A^n}{1-q_A^n-\tq^n},
\end{aligned} 
\end{equation}
which reduces to \eqref{c_n} when $q_A=0$.
The argument in \cite{Jafferis:2022wez} for the appearance of 
the geometric series 
is based on a diagramatic expansion of the two-matrix
model \eqref{eq:cZAB}. 
In our case, we have derived $c_n$ in \eqref{c_n}
by directly solving the saddle-point equation 
\eqref{G saddle} for $G(X,Y)$.

Note that the result \eqref{c_n} is consistent with the fact that 
our matrix model at $S=1$ reduces to the Gaussian matrix model $V(A)=\hf A^2$
in the limit $\tilde{q}\to0$ (see \eqref{eq:tq0-lim}). 
The two-point function of the Gaussian one-matrix model is 
\cite{Ambjorn:1990ji,Brezin:1993qg,Saad:2019lba} 
\begin{equation}
\left\langle {\rm Tr}\frac1{E_1-A}{\rm Tr}\frac1{E_2-A} \right\rangle_{\text{conn}}
=\frac1{2(E_1-E_2)^2}\left( \frac{E_1E_2-4}{\sqrt{(E_1^2-4)(E_2^2-4)}}-1 \right). 
\end{equation}
In order to compare this with our results, we set 
\begin{equation}
E_1=X+X^{-1}, \hspace{5mm} E_2=Y+Y^{-1}. 
\end{equation}
Suppose that $X,Y$ are small. 
If we choose the branch such that 
\begin{equation}
\sqrt{E_1^2-4}=X-X^{-1}, \hspace{5mm} \sqrt{E_2^2-4}=Y-Y^{-1}, 
\end{equation}
then we find 
\begin{equation}
\left\langle {\rm Tr}\frac{X-X^{-1}}{E_1-A}{\rm Tr}\frac{Y-Y^{-1}}{E_2-A} \right\rangle_{\text{conn}}=\frac{XY}{(1-XY)^2}=\sum_{n=1}^\infty nX^nY^n. 
\end{equation}
This coincides with our result \eqref{c_n} with $\tilde{q}=0$. 
On the other hand, if $X$ is small but $Y$ is large, then we should choose the other branch for $Y$ such that 
\begin{equation}
\sqrt{E_2^2-4}=-(Y-Y^{-1}). 
\end{equation}
Indeed, this choice reproduces the correct expansion 
\begin{equation}
\left\langle {\rm Tr}\frac{X-X^{-1}}{E_1-A}{\rm Tr}\frac{Y-Y^{-1}}{E_2-A} \right\rangle_{\text{conn}}=-\frac{XY}{(X-Y)^2}=-\sum_{n=1}^\infty nX^nY^{-n}
\end{equation}
for small $X$ and large $Y$. 

As another consistency check, let us consider the even part of $G(X,Y)$. 
One can show that the infinite sum in \eqref{infinite sum} can be recast 
into the form of the Bergman kernel 
\begin{equation}
\begin{aligned}
 G(X,Y) 
&= X\partial_XY\partial_Y\log\frac{\vartheta_1\left( \frac{\log XY^{-1}}{2\pi\mathrm{i}},\tilde{q} \right)}{\vartheta_1\left( \frac{\log XY}{2\pi\mathrm{i}},\tilde{q} \right)} \\
&= X\partial_XY\partial_Y\log\left[ \frac{X-Y}{XY-1}\prod_{n=1}^\infty \frac{(1-XY^{-1}\tilde{q}^n)(1-X^{-1}Y\tilde{q}^n)}{(1-XY\tilde{q}^n)(1-X^{-1}Y^{-1}\tilde{q}^n)} \right]. 
\label{eq:GXY-th}
\end{aligned} 
\end{equation}
The even part of \eqref{eq:GXY-th} then becomes 
\begin{eqnarray}
G_+(X,Y) 
&=& \frac12X\partial_XY\partial_Y\log\left[ \frac{X^2-Y^2}{X^2Y^2-1}\prod_{n=1}^\infty\frac{(1-X^2Y^{-2}\tilde{q}^{2n})(1-X^{-2}Y^2\tilde{q}^{2n})}{(1-X^2Y^2\tilde{q}^{2n})(1-X^{-2}Y^{-2}\tilde{q}^{2n})} \right] \nonumber \\
&=& \frac12X\partial_XY\partial_Y\log\frac{\vartheta_1\left( \frac{\log XY^{-1}}{\pi\mathrm{i}},\tilde{q}^2 \right)}{\vartheta_1\left( \frac{\log XY}{\pi\mathrm{i}},\tilde{q}^2 \right)}.
\label{eq:G+S1th} 
\end{eqnarray}
Using the relations for $S=1$
which we found in subsection \ref{sec:one-S1}
\begin{equation}
q=\tilde{q}^2, \hspace{5mm} X(u)=e^{\pi\mathrm{i}u},
\end{equation}
we can see that the even part
\eqref{eq:G+S1th}
of our solution
of $G(X,Y)$ agrees with $G_+(X,Y)$ in \eqref{eq:G+-bergman} when $S=1$.

\section{Conclusion and outlook}\label{sec:conclusion}
In this paper we have studied the
one- and two-point function of resolvents $\Tr\frac{1}{\la-A}$
in the two-matrix model for DSSYK in the limit $q_A,q_B\to0$.
In this limit, the matrix model potential
is given by the Gaussian terms plus the $q$-commutator squared interaction
\eqref{eq:VAB}.
After integrating out the matter matrix $B$, the partition function
of two-matrix model is written as the eigenvalue integral for the matrix $A$
and the large $N$ saddle-point equation can be 
solved in terms of an elliptic function.
To study this model, 
it is convenient to introduce 
the 't Hooft parameter $S$ in the potential
\eqref{eq:VAB-S}.
It turned out that the solution of planar resolvent behaves
differently for $S<1$ and $S=1$.
When $S=1$, we confirmed that the moments of two-point function
in \cite{Jafferis:2022wez} are correctly reproduced from
our planar solution
of $G(X,Y)$ in \eqref{infinite sum}.

There are many interesting open questions.
We would like to understand the bulk Hilbert space of
quantum gravity coupled to matter fields.
One can show that the moments 
$c_{n,m}$ \eqref{eq:cnm} of the two-point function at $S=1$ is written as
(see e.g. appendix B of \cite{Zhou:2023eus})
\begin{equation}
\begin{aligned}
c_{n,m}
=\frac{\Tr (\tq^{L_0}\al_n\al_{-m})}{\Tr \tq^{L_0}}=\frac{n}{1-\tq^n}\cob_{n,m},
\end{aligned}
\label{eq:moment-boson} 
\end{equation}  
where $\al_n$ is the free boson with the commutator
$[\al_n,\al_m]=n\cob_{n+m,0}$ and
$L_0$ is the Virasoro generator
\begin{equation}
\begin{aligned}
 L_0=\hf\al_0^2+\sum_{n=1}^\infty \al_{-n}\al_n.
\end{aligned} 
\end{equation}
This expression \eqref{eq:moment-boson} suggests that
the bulk Hilbert space of
the wormhole geometry is isomorphic to the Fock space of
free boson.
This is not so surprising since it is known that
the matrix model correlators admit a CFT representation \cite{Kostov:2009nj,Kostov:2010nw}.
In this computation \eqref{eq:moment-boson}, the zero mode $\al_0$ is decoupled
and it does not contribute to the two-point function.
It is tempting to identify the zero mode $\al_0$ as the
$\mathcal{L}^2(\mathbb{R})$ factor of the Hilbert space 
for the wormhole geometry discussed in \cite{Chen:2023hra}.
Note that the eigenvalue of $L_0$ corresponds
to the discrete length of the bulk geodesic loop
running around the neck of the wormhole 
\cite{Jafferis:2022wez,Okuyama:2023byh,Okuyama:2023kdo}.
The $L_0=0$ state corresponds to a thin wormhole, 
but such a state is decoupled in the computation of 
two-point function (i.e. there is no $n=0$ term in
\eqref{eq:om-cn}).
Thus the two-point function in our two-matrix model
is UV finite.
It would be interesting to understand the bulk Hilbert space
of our model better.

It would also be interesting to study the ETH matrix model of DSSYK
when $q_A,q_B\ne0$
and compute the two-point function of resolvents.
Perhaps, the Virasoro generator $L_0$
in \eqref{eq:moment-boson} might be replaced by the $q$-Virasoro generator
\cite{Shiraishi:1995rp} when $q_A,q_B\ne0$.
We leave this generalization as an interesting future problem.

\acknowledgments
The work of KO was supported
in part by JSPS Grant-in-Aid for Transformative Research Areas (A) 
``Extreme Universe'' 21H05187 and JSPS KAKENHI Grant 22K03594.

\appendix
\section{Even moments of two-point function for $S<1$}\label{app:moment}
From the result of $G_+(X,Y)$ in \eqref{G_+}, we can obtain the
even part of the moments $d_{n,m}$ defined by 
\begin{equation}
\lim_{u_X,u_Y\to u_0}2\frac\partial{\partial \log X^2}\frac\partial{\partial \log Y^2}\log\frac{1-X^2Y^2}{\vartheta_1(u_X+u_Y)}=2\sum_{n,m=1}^\infty d_{n,m}X^{2n}Y^{2m}. 
\end{equation}
For instance, $d_{1,1}$ is given by 
\begin{eqnarray}
d_{1,1}
&=& \frac1{\vartheta_1'(0)^2}\left[ \vartheta_1'(2u_0)^2-\vartheta_1''(2u_0)\vartheta_1(2u_0) \right]-1 \nonumber \\
&=& \frac{1-\tilde{q}^2}{\tilde{q}^2}+(S-1)\left[ \tilde{q}^{-2}+S^3-1+3(S-1)S^3(4S-1)\tilde{q}^2+{\cal O}(\tilde{q}^4) \right]. 
\end{eqnarray}
In general, the off-diagonal term $d_{n,m}\,(n\ne m)$ is non-zero when $S<1$.
However, it turns out that $d_{n,m}$ becomes diagonal in 
the $S\to1$ limit
\begin{equation}
\lim_{S\to1}d_{n,m}=n\frac{1-\tilde{q}^{2n}}{\tilde{q}^{2n}}\delta_{n,m}. 
   \label{lim moment}
\end{equation}
We have checked this relation for $(n,m)=(1,1),(2,1),(2,2)$. 
Note that the right-hand side of \eqref{lim moment} is obtained from $c_{2n}$ in 
\eqref{c_n} as 
\begin{eqnarray}
T_XT_Y\sum_{n=1}^\infty c_{2n}X^{2n}Y^{2n} 
&=& 2\sum_{n=1}^\infty\frac n{1-\tilde{q}^{2n}}\frac{(1-\tilde{q}^{2n})^2}{\tilde{q}^{2n}}X^{2n}Y^{2n} \nonumber \\
&=& 2\sum_{n=1}^\infty n\frac{1-\tilde{q}^{2n}}{\tilde{q}^{2n}}X^{2n}Y^{2n}. 
\end{eqnarray}
This computation is different from the one we performed in deriving $c_n$
in \eqref{eq:TXTY-S1}. 
This difference comes from the fact that $X(u)$
in \eqref{eq:Xu} vanishes at $u=u_0$ for $S<1$ 
while $X(u)=e^{\pi\mathrm{i}u}$ for $S=1$ does not vanish at $u=u_0$. 
\bibliography{paper}
\bibliographystyle{utphys}

\end{document}